\documentclass[fleqn,usenatbib]{mnras}

\usepackage{newtxtext,newtxmath}

\usepackage[T1]{fontenc}
\usepackage{ae,aecompl}

\usepackage{graphicx}	
\usepackage{amsmath}	
\usepackage{amssymb}	

\title[Variable HFQPO in GRS 1915+105]{A variable-frequency HFQPO in GRS 1915+105 as observed with \textsc{Astrosat}}

\author[T.M. Belloni et al.]{
Tomaso M. Belloni,$^{1}$\thanks{E-mail: tomaso.belloni@inaf.it}
Dipankar Bhattacharya,$^{2}$
Pietro Caccese,$^{3}$
Varun Bhalerao,$^{4}$
\newauthor Santosh Vadawale,$^{5}$
and J.S. Yadav$^{6}$
\\
$^{1}$INAF - Osservatorio Astronomico di Brera, via E. Bianchi 46, I-23807 Merate, Italy\\
$^{2}$Inter University Centre for Astronomy and Astrophysics, Post Bag 4, Pune 411007, India\\
$^{3}$Liceo Scientifico ``Giuseppe Mercalli," via Andrea d'Isernia, 34, 80122, Napoli, Italy\\
$^{4}$Department of Physics, Indian Institute of Technology Bombay, Powai, Mumbai-400076, India\\
$^{5}$Physical Research Laboratory, Navrangpura, Ahmadabad 380009, India\\
$^{6}$Tata Institute of Fundamental Research, 400005, Mumbai, Maharashtra, India
}

\date{Accepted XXX. Received YYY; in original form ZZZ}

\pubyear{2019}

\begin{document}
\label{firstpage}
\pagerange{\pageref{firstpage}--\pageref{lastpage}}
\maketitle

\begin{abstract}
From the analysis of more than 92 ks of data obtained with the laxpc instrument on board \textsc{Astrosat} we have detected a clear high-frequency QPO whose frequency varies between 67.4 and 72.3 Hz. In the classification of variability classes of GRS 1915+105, at the start of the observation period the source was in class $\omega$ and at the end the variability was that of class $\mu$: both classes are characterized by the absence of hard intervals and correspond to disk-dominated spectra. After normalization to take into account time variations of the spectral properties as measured by X-ray hardness, the QPO centroid frequency is observed to vary along the hardness-intensity diagram, increasing with hardness. We also measure phase lags that indicate that HFQPO variability at high energies lags that at lower energies and detect systematic variations with the position on the hardness-intensity diagram.
This is the first time that (small) variations of the HFQPO frequency and lags are observed to correlate with other properties of the source. We discuss the results in the framework of existing models, although the small (7\%) variability observed is too small to draw firm conclusions.
\end{abstract}




\section{Introduction}
\label{sec:introduction}

The study of fast time variability of X-ray emission from black-hole binaries (BHB) has received a considerable boost during the sixteen years of operation of the Rossi X-ray Timing Explorer (RXTE) mission, that has provided millions of seconds of high-sensitive observations of both transient and persistent systems. In addition to broad-band noise components and Low-Frequency Quasi-Periodic Oscillations (LFQPOs, in the frequency range 0.01-30 Hz), which had already been observed in the past with previous missions such as EXOSAT and Ginga, RXTE discovered High-Frequency QPOs (HFQPOs) at frequencies above 30 Hz, with a current highest measured frequency of 450 Hz (in GRO J1655-40, see Remillard et al. 1999). 
HFQPOs are important as their frequency is in the range expected for Keplerian motion of matter in the vicinity of the black hole and can be a direct way of 
exploring space-time near a collapsed object. Unfortunately, to date we have very few detections of them, obtained with RXTE and all corresponding to intervals when source fluxes were very high. This could be either because the signal is present only during those high-flux states and/or that a high count rate is needed to reach a significant detection (see Belloni, Sanna \& M\'endez 2012 and references therein). Moreover, a clear identification with a physical time scale in the accretion flow, whether associated to accretion properties or to General Relativity, can only come from the detection of multiple frequencies. While in very few cases double HFQPO peaks have been detected, these features appear to be visible only when LFQPOs are not detected. The only two cases of multiple detections have been analyzed by Motta et al. (2014a,b) and identified with the Relativistic Frequencies predicted by the Relativistic-Precession Model (RPM, Stella \& Vietri 1998,1999), leading to a measurement of mass and spin of the black hole in one case and of the spin in the other. 

An important exception to the parsimoniousness of black-hole binaries with HFQPOs is the bright source GRS 1915+105. The source appeared as a very bright transient in 1992 and since then it has remained bright (see Fender \& Belloni 2004 for a review). In addition to its high flux, which can reach the Eddington level, GRS 1915+105 displays very peculiar variability on time scales longer than a second, with structured patterns that repeat even after many years (see Belloni et al. 2000; Belloni 2010 for reviews). These variations, which have been associated to disk instabilities (see e.g. Belloni et al. 1997a,b; Janiuk, Czerny \& Siemiginowska 2000), involve major spectral and intensity changes and have been classified into a dozen of separate classes (Belloni et al. 2000; Klein-Wolt et al. 2002; Hannikainen et al. 2005).
The first HFQPO was discovered in the early RXTE data of  GRS 1915+105 (Morgan, Remillard \& Greiner 1997) at a frequency of $\sim$65-67 Hz, with a low fractional rms around 1\%, which increased to $\sim$10\% at high energies. Transient additional high-frequency peaks have been discovered later in selected observations (see Belloni \& Altamirano 2013a and references therein). A systematic analysis of the full set of RXTE observations of GRS 1915+105, for a total of more than $5\times 10^6$ s exposure, was performed by Belloni \& Altamirano (2013a). From this work, a total of 51 HFQPO peaks were detected, most of which in the 65-70 Hz range.

The demise of the RXTE satellite left us without an instrument capable of efficiently detecting HFQPOs such as its Proportional Counter Array (PCA). The key necessary feature, in addition to high-time resolution, is a large collecting area at energies above a few keV, where HFQPOs are more intense. The launch of the \textsc{Astrosat} mission in September 2015 filled this gap. We report here the results of the analysis of a series of \textsc{Astrosat} observations of GRS 1915+105 made in July-September 2017, when a clear HFQPO was detected.

\section{Observations and data analysis}
\label{sec:observations}
\begin{figure}
	\includegraphics[width=\columnwidth]{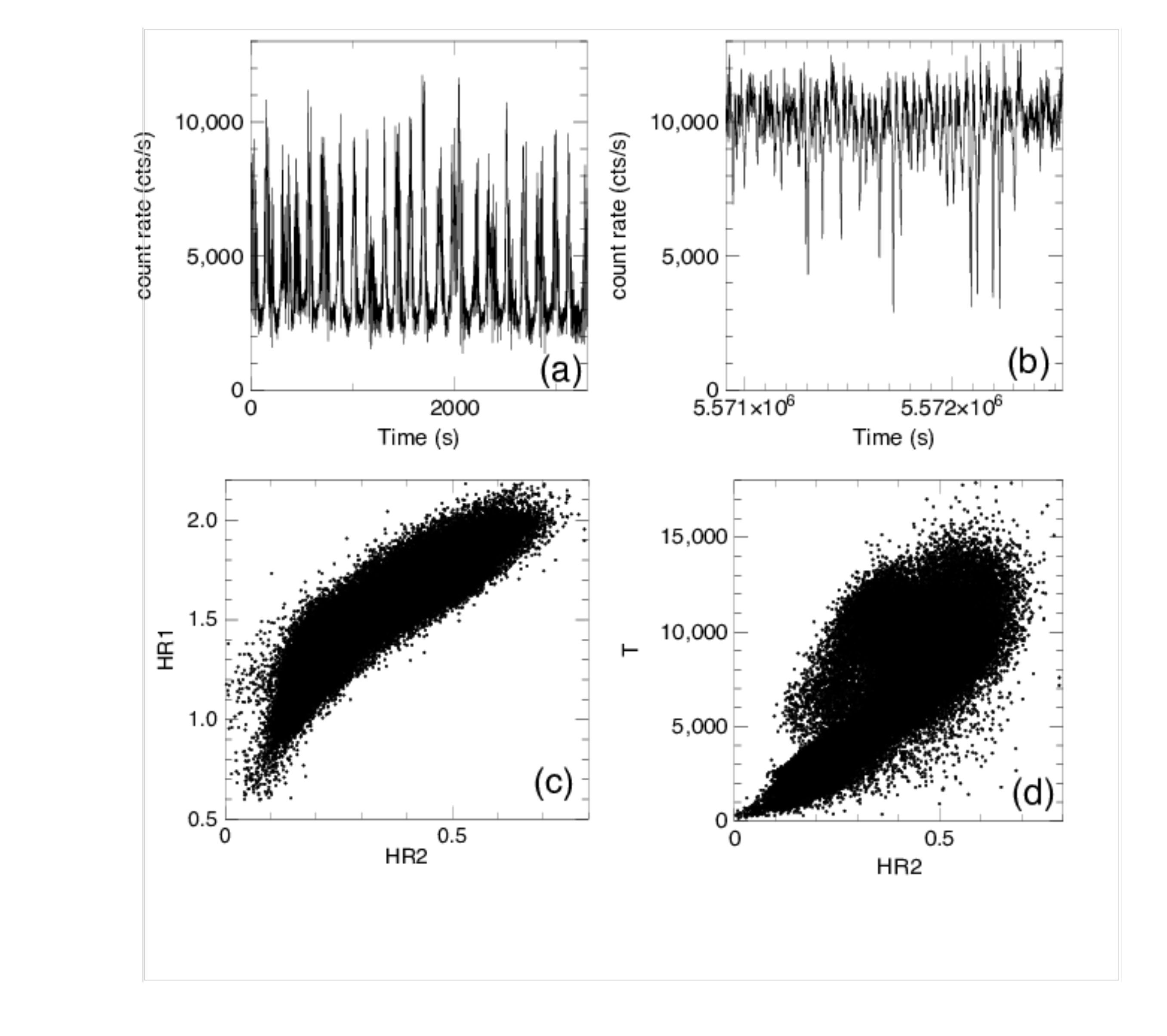}
    \caption{(a) Light curve of the start of the observation (bin size 1.024s, start time MJD 57943.71869).
             (b) Light curve of the end of the observation (bin size 1.024s, same start time).
             (c) Color-color diagram for the full observation, before the shift/rotation described below.
             (d) Hardness-intensity diagram for the full observation, before the shift/rotation described below.
             }
    \label{fig:licu}
\end{figure}

We analyzed a set of observations of GRS 1915+105 taken with the LAXPC instrument on board \textsc{Astrosat} (Singh et al. 2014). The data were obtained from the \textsc{Astrosat} public archive (https://astrobrowse.issdc.gov.in/astro\_archive/archive/Home.jsp).
The LAXPC is an X-ray proportional counter array operating in the range 3-80 keV. The timing resolution of the instrument is 10 $\mu$s with a dead time of 42 $\mu$s. It consists of three identical detectors (referred to as LX10, LX20 and LX30 respectively), with a combined effective area of 6000 cm$^2$ (Yadav et al. 2016; Antia et al. 2017). For the full observation, all information about single photons is available.
In order to obtain photon lists, we started from level1 production files from the archive and converted them to level2 using the LaxpcSoft tools provided by the \textsc{Astrosat} mission (see http://astrosat-ssc.iucaa.in). The tools provide a channel-to-energy conversion using the appropriate detector response matrices for the three instruments, minimizing effects due to gain changes. Therefore, the energy selection was not based on channels, but on the energies estimated with the tools. Net light curves, including those for the production of hardness ratios (see below) were obtained by subtracting the background estimated with the same tools. No deadtime correction was applied.
We analyzed six observations, listed in Tab. \ref{tab:observations}.
They cover a time period from 2017 July 9 16:46UT to 2017 September 12 04:54UT, for a total of 92.293 ks of net exposure, consisting of several satellite orbits. The observations were selected in order to have a similar behaviour of variability uninterrupted by observations of different type. Most of the observations see the source in its $\omega$ variability class (see Belloni et al. 2000, Klein-Wolt et al. 2002), although a secular evolution in the properties towards class $\mu$ is observed (see top panels in Fig. \ref{fig:licu}). 

\begin{table}
	\centering
	\caption{Log of the six data intervals in 2017. The last column reports the number of 1.024s data stretches contained in the interval.}
	\label{tab:observations}
	\begin{tabular}{lcc} 
		\hline
		Obs. ID                                 & T$_{start}$ & T$_{end}$\\
		\hline
		G07\_028T01\_9000001370 &July               9 16:46&July              9 22:47\\
                 G07\_046T01\_9000001374 &July             11 20:07&July            12 11:59\\
                 G07\_028T01\_9000001406 &July             27 09:21&July            27 14:13\\
                 G07\_046T01\_9000001408 &July             27 14:18&July            28 04:45\\
                 G07\_028T01\_9000001500 &August        30 02:15&August        30 09:54\\
                 G07\_046T01\_9000001534 &September  11 13:38&September 12 05:17\\
		\hline
	\end{tabular}
\end{table}

Using the GHATS analysis package, developed at INAF-OAB for the analysis of variability from X-ray datasets (http://www.brera.inaf.it/utenti/belloni/GHATS\_Package/Home.html) we extracted light curves with a 1.024s bin size for the energy bands 3-80 keV (band I), 3-5 keV (band A), 5-10 keV (band B) and 10-20 keV (band C), summing all three units and all instrument layers. No background subtraction was applied, as it would alter the statistical properties of the data without providing any advantage. From these we produced two X-ray hardness parameters HR1=B/A and HR2=C/A. The full color-color diagram (CCD: HR1 vs. HR2) has a very defined but broad shape, while the hardness-intensity diagram is even broader (HID: I vs. HR2), as can be seen in Fig. \ref{fig:licu}. Using GHATS, we extracted Power Density Spectra (PDS) from data stretches 1.024s long in the 3-80 keV band, corresponding to the times of the points in the diagrams, adding all detectors and all layers. The PDS were normalized after Leahy et al. (1983) and extended to a Nyquist frequency of 500 Hz. We then averaged the PDS in small regions of the HID and searched for HFQPOs in the 30-200 Hz band. Clear peaks around 70 Hz were seen in the different HID regions with variable frequency, but no systematic variations could be determined and broad and multiple peaks were seen. In order to ascertain whether these variations were due to secular variations in the HID shape, we produced the HID corresponding to six time intervals separated by large time gaps in the data. Intervals E and F are closer in time, but as there are differences in the time evolution we decided to keep them separate. The time limits of the intervals are shown in Tab. \ref{tab:intervals}.

\begin{table}
	\centering
	\caption{Log of the six data intervals in 2017. The last column reports the number of 1.024s data stretches contained in the interval.}
	\label{tab:intervals}
	\begin{tabular}{lccc} 
		\hline
		Interval & T$_{start}$ & T$_{end}$&N$_{PDS}$\\
		\hline
		A & July       9 16:46   &  July      9 22:17 & 9639  \\
		B & July      11 19:39   &  July     12 11:28 & 18430 \\
		C & July      27 08:39   &  July     28 04:16 & 28601 \\
		D & August    30 02:02   & August    30 09:24 & 9903  \\
		E & September 11 13:37   & September 11 17:25 & 6722  \\
		F & September 11 18:54   & September 12 04:54 & 11592 \\
		\hline
	\end{tabular}
\end{table}

\begin{figure}
	\includegraphics[width=\columnwidth]{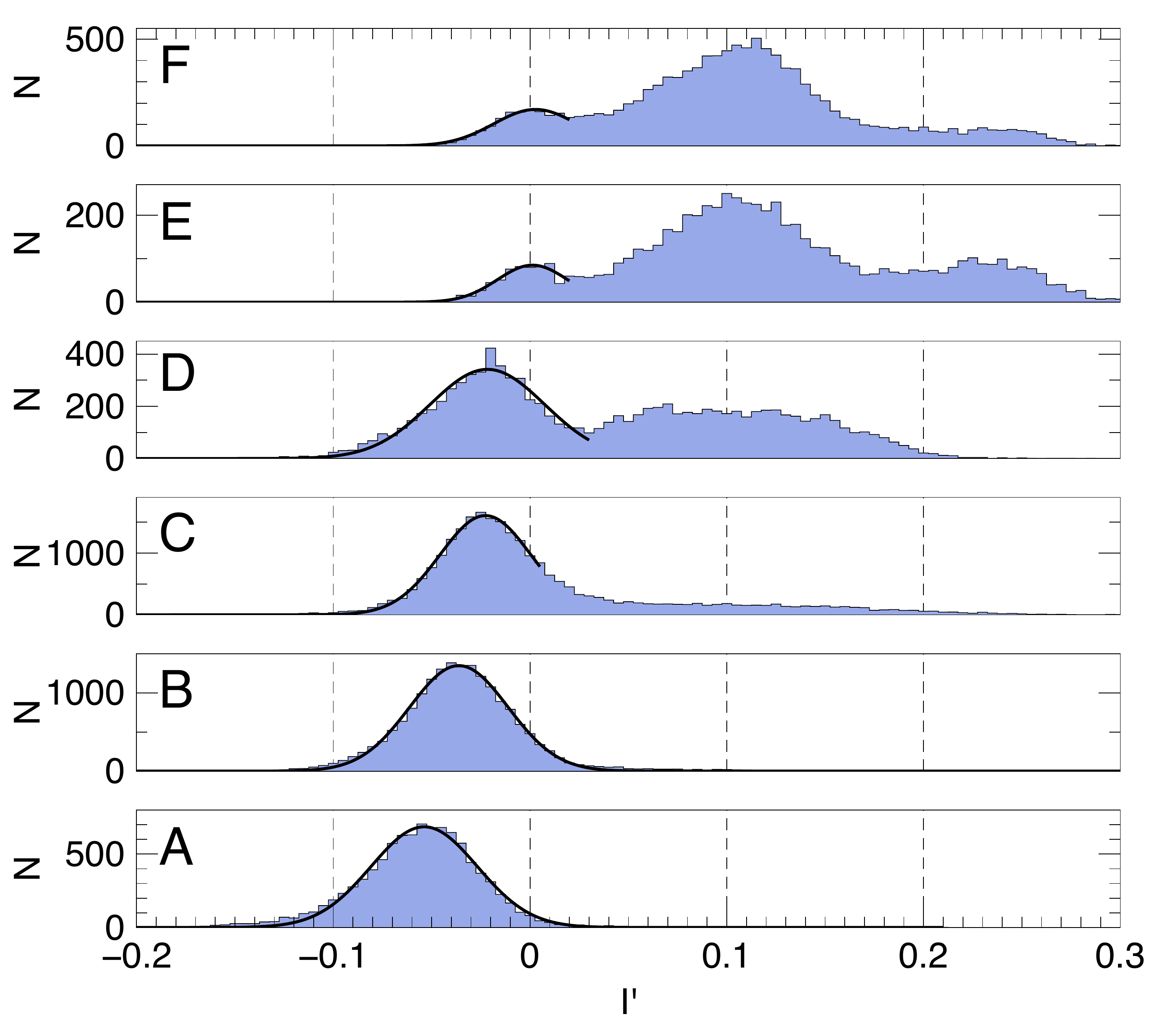}
    \caption{Marginal distributions in $I'$ for the six intervals defined in Tab. \ref{tab:intervals}. The black lines are the best-fit Gaussian models to the leftmost peak, plotted over the range used for each fit.
             }
    \label{fig:marginals}
\end{figure}

The HID of the six intervals appear shifted. In order to ascertain the best shift values, noticing that the main direction of the HID 2D distribution is diagonal, we renormalised the count rate $I$ as $J=I/17500$ and rotated the HID counter-clockwise by 45 degrees obtaining two new coordinates $H'$ and $I'$. The resulting H$'$I$'$D has the shape of a mirrored L. With this rotation, it appeared clear that the shift between different intervals was only in the vertical direction $I'$. The marginal distribution in $I'$ for the six intervals are shown in Fig. \ref{fig:marginals}.

\begin{figure}
	\includegraphics[width=\columnwidth]{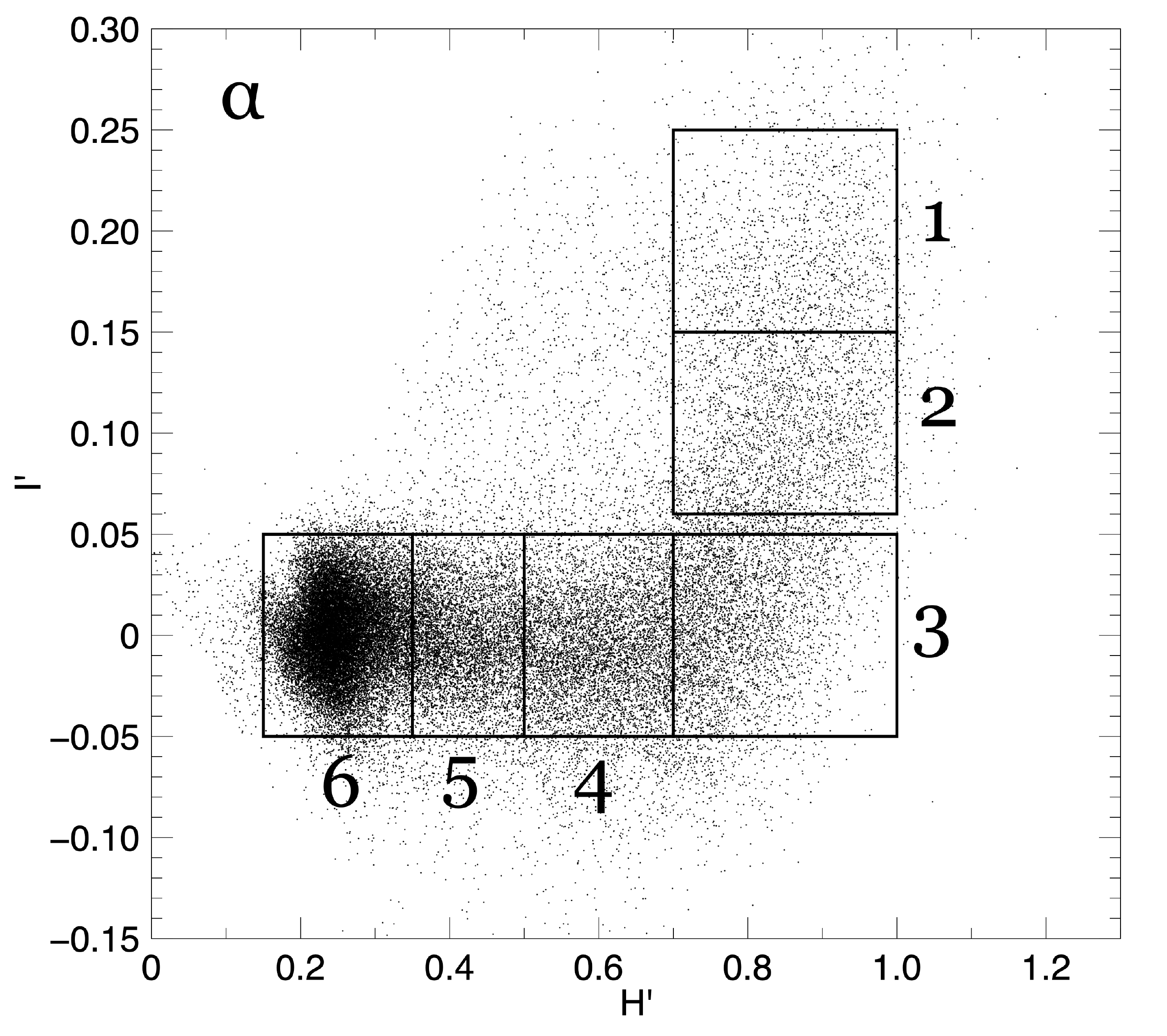}
	\includegraphics[width=\columnwidth]{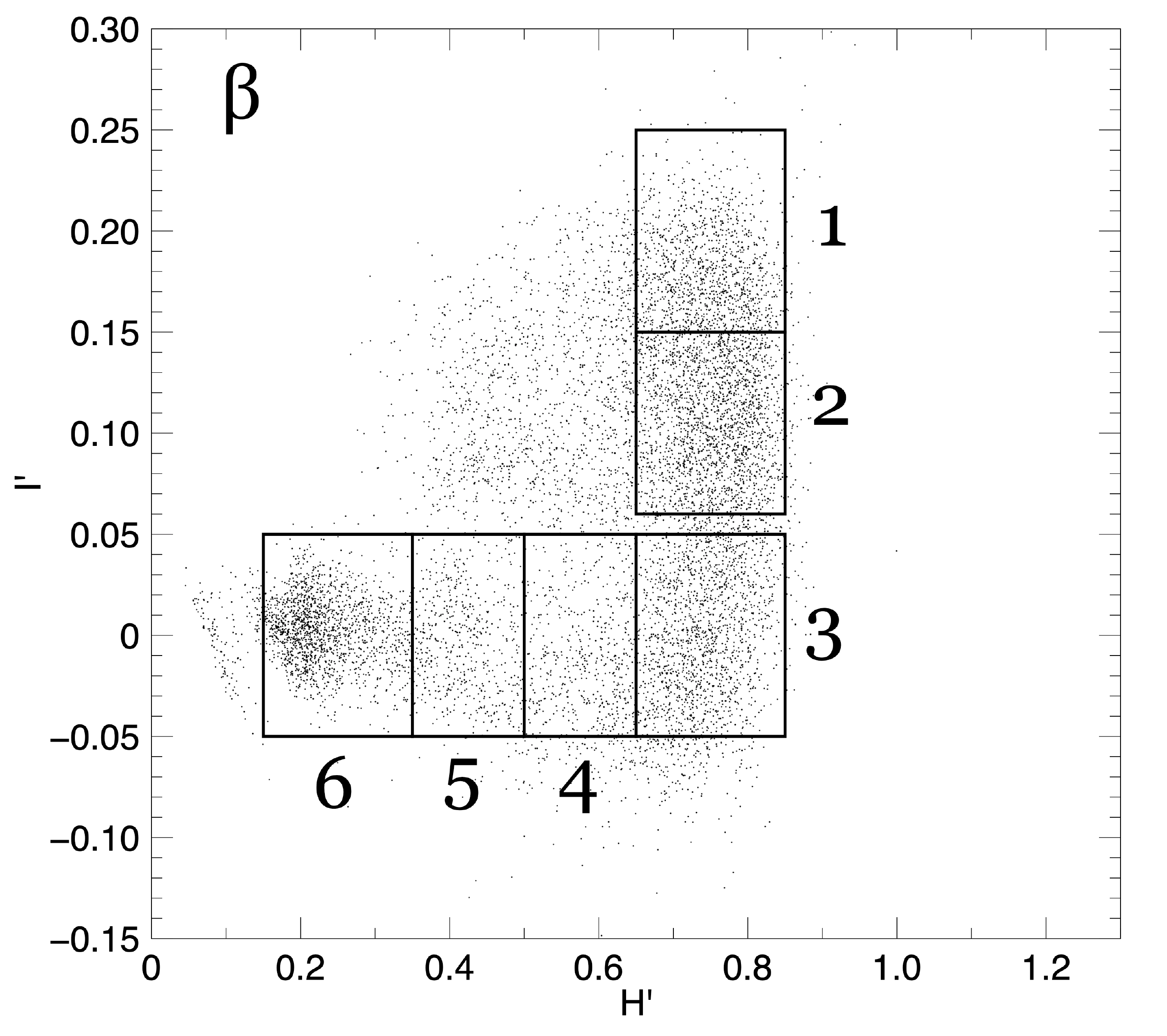}
	\includegraphics[width=\columnwidth]{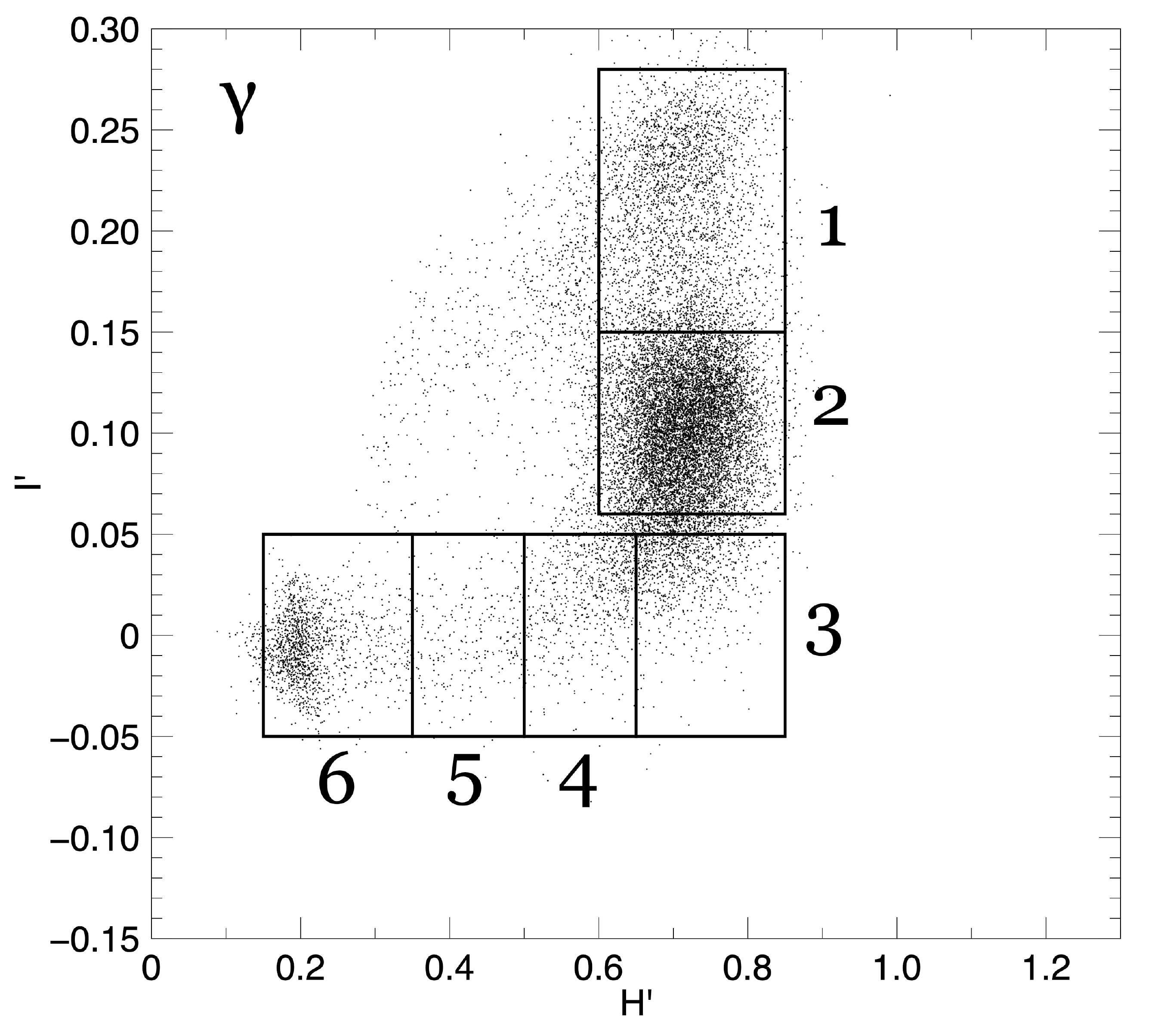}
    \caption{The three H$'$I$'$Ds for the three groups (see text). The extraction regions 1 through 6 for each group are shown.
             }
    \label{fig:hid}
\end{figure}

We fitted a Gaussian function to the leftmost peak in the distributions in Fig. \ref{fig:marginals}, limiting the range in abscissa in order to obtain a good fit without the need to include other peaks (see Fig. \ref{fig:marginals}). We then shifted the H$'$I$'$Ds for the six intervals to coalign the peak of the Gaussians. Upon visual inspection, there were still differences between the shifted H$'$I$'$D, therefore we grouped the six intervals into three groups: $\alpha$ (A,B,C), $\beta$ (D) and $\gamma$ (E,F), obtaining three much more well-defined H$'$I$'$Ds distributions, which can be seen in Fig. \ref{fig:hid}.

For each of the three groups, we identified six regions to cover the main part of the H$'$I$'$D. The regions can be seen in Fig. \ref{fig:hid}. They were chosen in order to cover the track, include the bulk of the points, but leave stragglers out as they are rather distant from the main track. Since the valley between the peaks in regions 2 and 3 are not very deep (see Fig. \ref{fig:marginals}), we left a gap between these two regions.


\section{Results}

For each of the six regions in each of the three groups, we averaged the PDS, obtaining 18 final PDS. We fitted each PDS in the 30-200 Hz region with a model consisting of a power law (to account for Poissonian noise) and a Lorentzian peak.

\begin{figure}
	\includegraphics[width=\columnwidth]{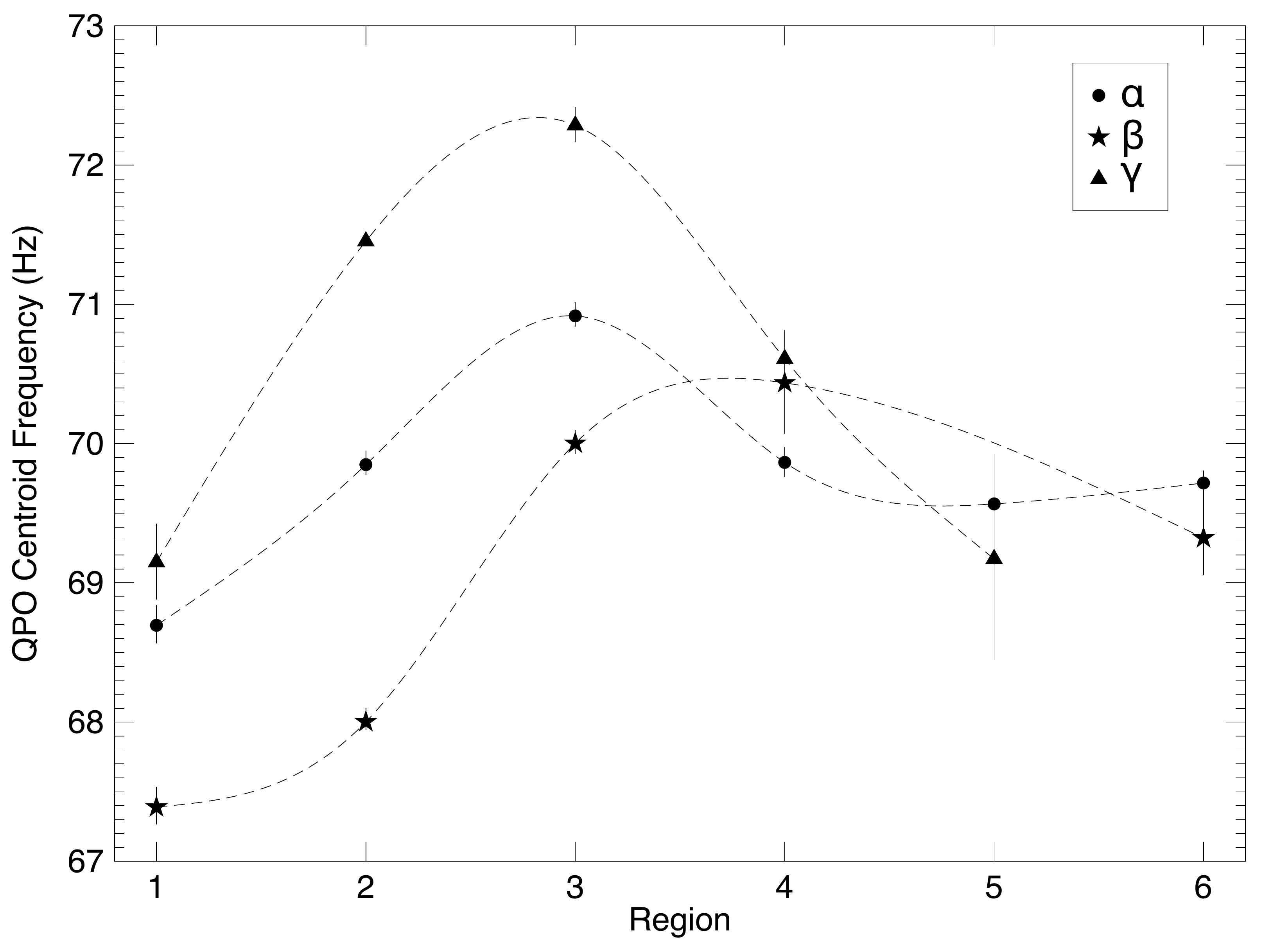}
    \caption{Centroid frequency of the HFQPO as a function of H$'$I$'$Ds region for the three selection groups. All frequencies have an error bar, often within the symbol.}
    \label{fig:centroid}
\end{figure}

\begin{figure}
	\includegraphics[width=\columnwidth]{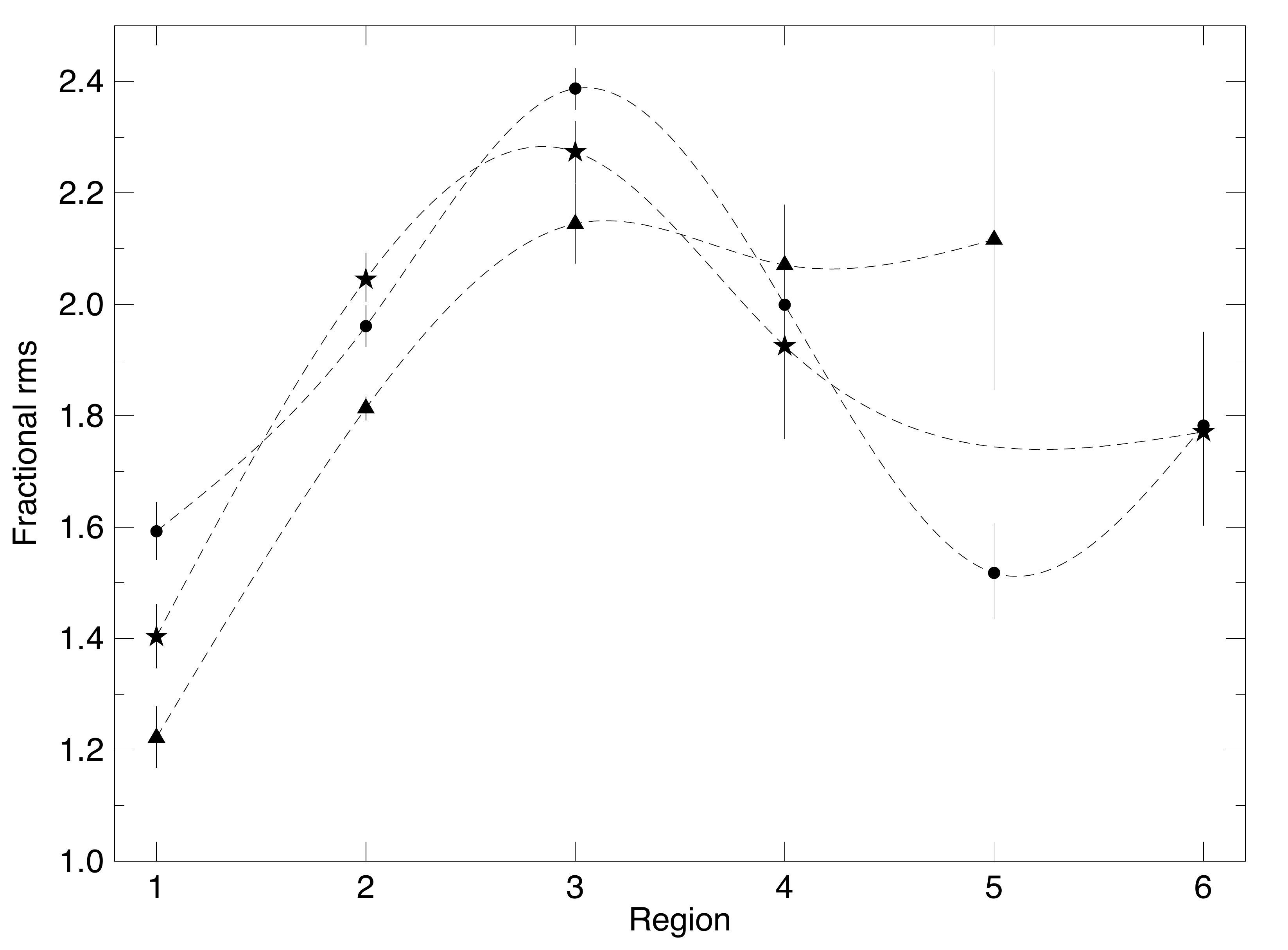}
    \caption{Integrated fractional rms of the HFQPO as a function of H$'$I$'$Ds region for the three selection groups.}
    \label{fig:rms}
\end{figure}

\begin{figure}
	\includegraphics[width=\columnwidth]{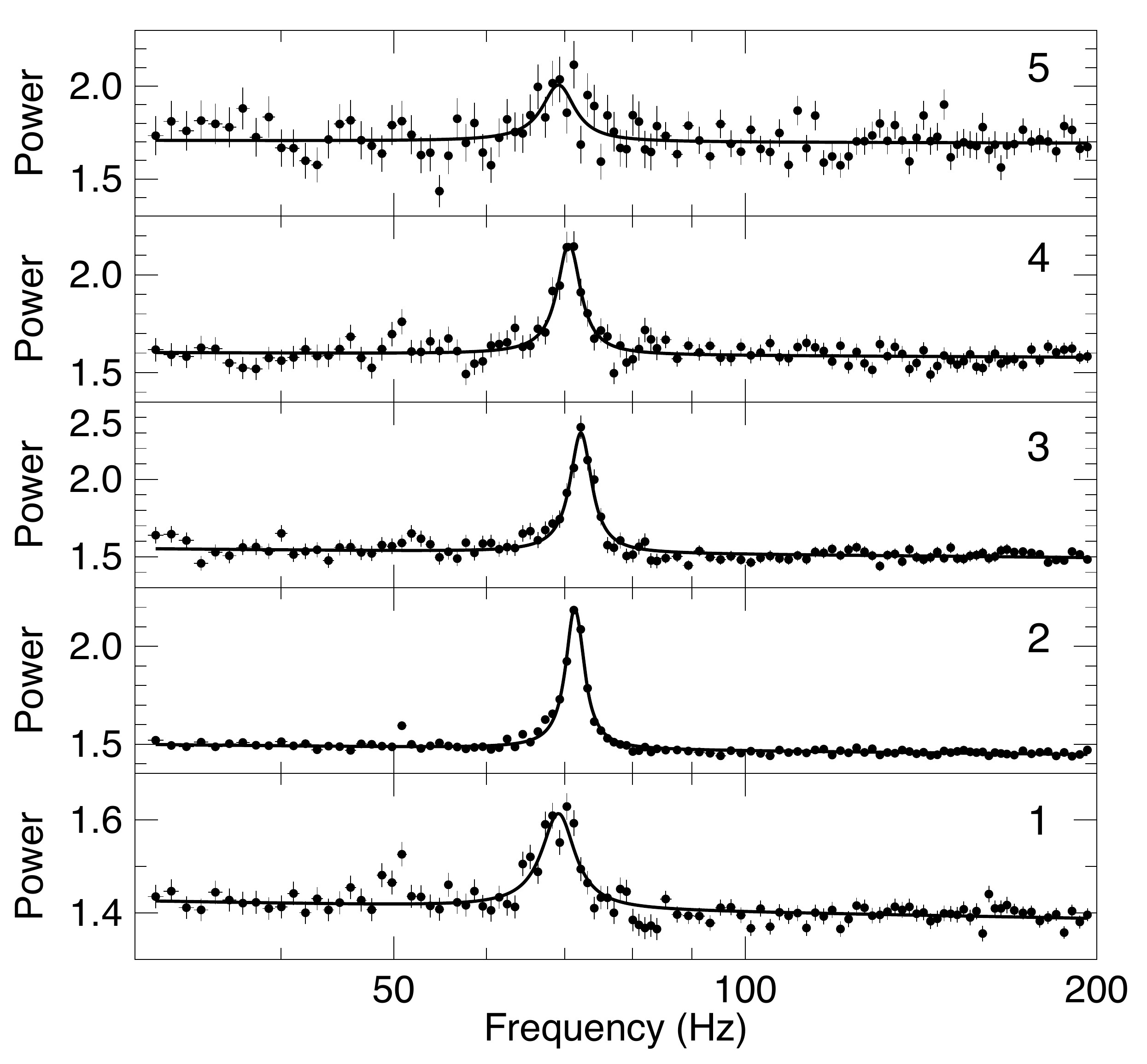}
    \caption{PDS from group $\gamma$ together with their best fit. The changes in centroid frequency are evident.}
    \label{fig:pds}
\end{figure}

A significant (more than 3$\sigma$) detection of a $\sim$ 70 Hz QPO peak is found for all eighteen PDS, with the exception of two ($\beta5$ nd $\gamma6$). The QPO parameters are shown in Tab. \ref{tab:qpo}. For all three groups, the centroid frequency of the QPO is not constant and shows the same evolution as a function of segment number, shown in Fig. \ref{fig:centroid}. The frequency increases from region 1, peaks at region 3 (region 4 for $\beta$), then stabilizes around 69.5 Hz for all three groups, overall varying between 67.4 Hz and 72.3 Hz. The quality factor (defined as the ratio of the centroid frequency to the FWHM of the Lorentzian peak) is between 12 and 28, without clear trends. The integrated fractional rms of the QPO peak (in the 3-80 keV energy band) is shown in Fig. \ref{fig:rms}. As an example, the PDS for group $\gamma$ are plotted in Fig. \ref{fig:pds}, where the changes in centroid frequency are evident.

\begin{figure}
	\includegraphics[width=\columnwidth]{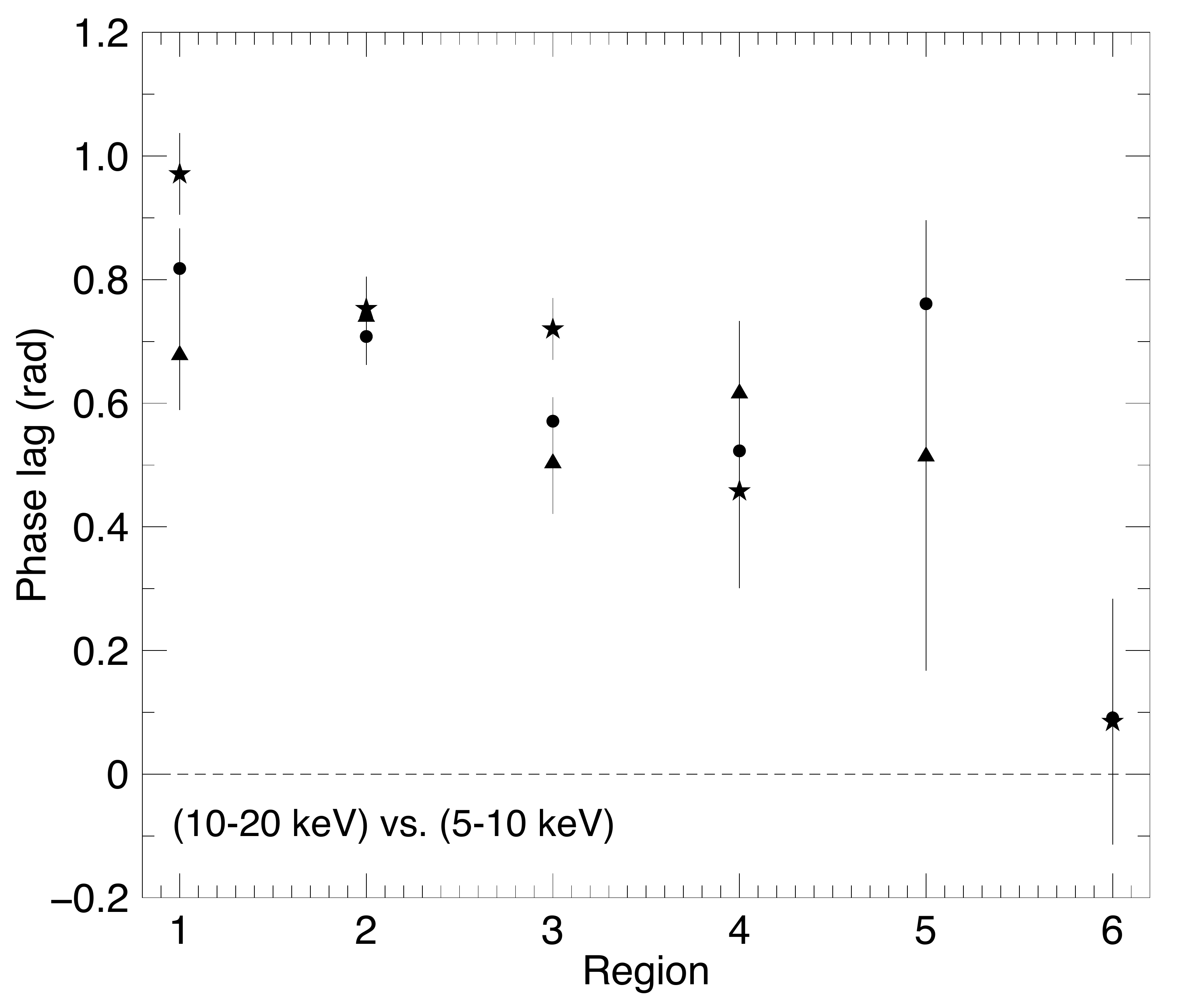}
	\includegraphics[width=\columnwidth]{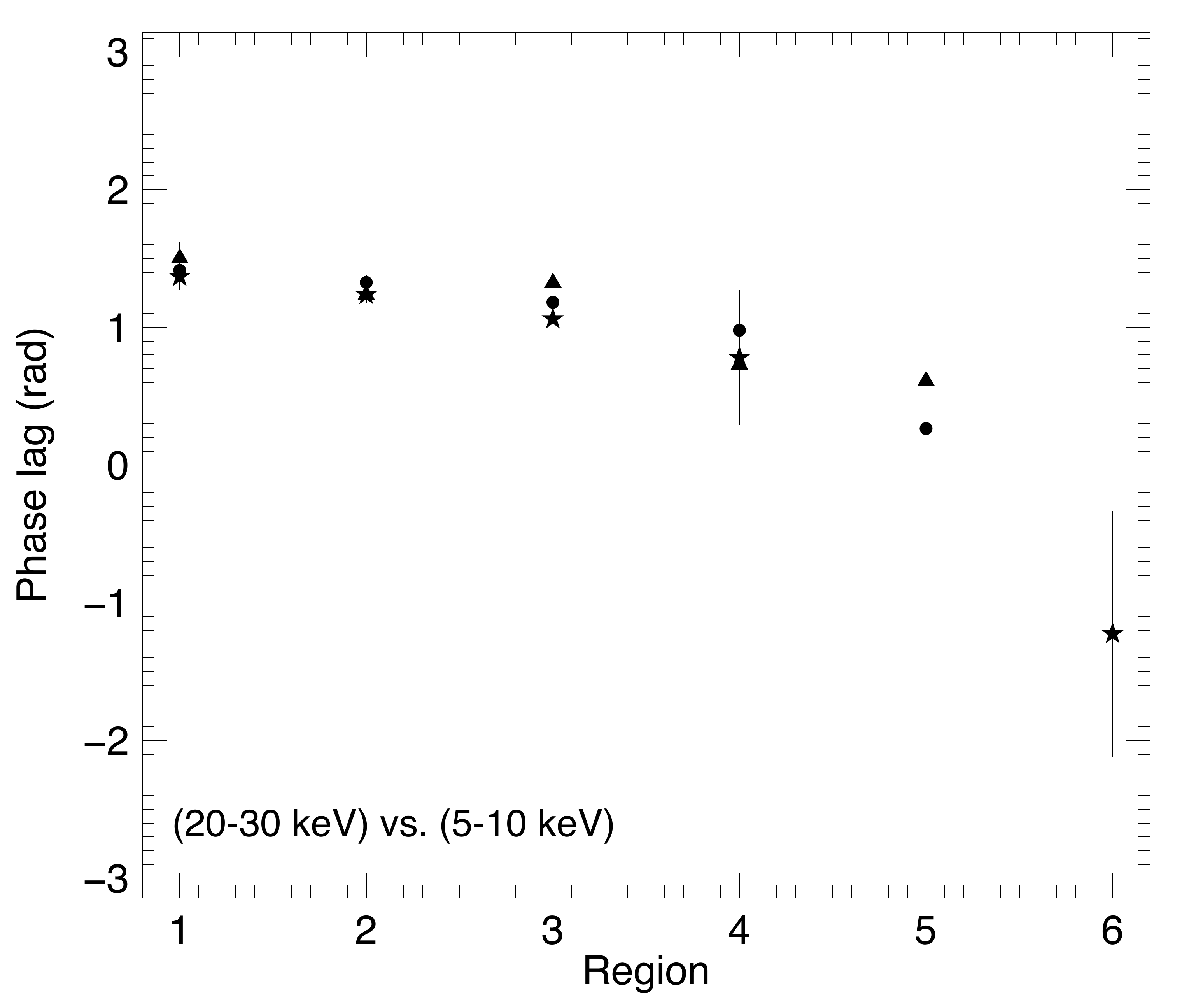}
    \caption{Phase lags at the HFQPO as a function of H$'$I$'$Ds region for the three selection groups. All frequencies have an error bar, often within the symbol. Top: 10-20 keV vs. 5-10 keV. Bottom: 20-30 keV vs. 5-10 keV.
             }
    \label{fig:lags}
\end{figure}

We produced cross spectra over the same 1.024s stretches used for the PDS between the counts in the 5-10 keV energy band and those in the 10-20 keV and 20-30 keV bands, leaving the 3-5 keV band as the signal there is too faint (low rms and low count rate) to yield significant results. We averaged the cross spectra over the same eighteen regions as the PDS. From each averaged cross spectrum we averaged the complex values in a frequency range $\nu_0-\Delta/2$ -- $\nu_0+\Delta/2$ and from the average calculated the phase lag at the QPO (see M\'endez et al. 2013). In order to account for cross-channel talk, we subtracted from the complex value an average value computed in the 110-190 Hz band, where only Poissonian noise is present in the PDS.  The phase lags as a function of region for the three groups are shown in Fig. \ref{fig:lags}. For both energy ranges, the phase lags are positive (hard lags soft) and decrease with region number. In the 20-30 keV case region 6 reaches a negative (soft lags hard).

\begin{table}
	\centering
	\caption{QPO parameters for the regions shown in Fig. \ref{fig:hid}}
	\label{tab:qpo}
	\begin{tabular}{lccc} %
		\hline
		Region & $\nu_0$ (Hz)       & FWHM (Hz)      & \% rms\\
		\hline
		$\alpha$1 & 68.70 $\pm$ 0.14& 3.84 $\pm$ 0.35& 1.59 $\pm$ 0.05\\
		$\alpha$2 & 69.85 $\pm$ 0.10& 4.15 $\pm$ 0.22& 1.96 $\pm$ 0.04\\
		$\alpha$3 & 70.92 $\pm$ 0.09& 5.33 $\pm$ 0.24& 2.39 $\pm$ 0.04\\
		$\alpha$4 & 69.87 $\pm$ 0.10& 4.65 $\pm$ 0.26& 2.00 $\pm$ 0.04\\
		$\alpha$5 & 69.57 $\pm$ 0.29& 5.11 $\pm$ 0.70& 1.52 $\pm$ 0.08\\
		$\alpha$6 & 69.72 $\pm$ 0.12& 3.41 $\pm$ 0.25& 1.78 $\pm$ 0.05\\
		\hline
		$\beta$1  & 67.39 $\pm$ 0.14& 2.45 $\pm$ 0.26& 1.40 $\pm$ 0.06\\
		$\beta$2  & 68.00 $\pm$ 0.10& 3.08 $\pm$ 0.20& 2.04 $\pm$ 0.04\\
		$\beta$3  & 70.00 $\pm$ 0.10& 3.20 $\pm$ 0.23& 2.27 $\pm$ 0.06\\
		$\beta$4  & 70.44 $\pm$ 0.37& 4.38 $\pm$ 1.24& 1.93 $\pm$ 0.17\\
		$\beta$5  & -               & -              & -              \\
		$\beta$6  & 69.32 $\pm$ 0.35& 2.53 $\pm$ 0.77& 1.77 $\pm$ 0.18\\
		\hline
		$\gamma$1 & 69.15 $\pm$ 0.27& 5.42 $\pm$ 0.62& 1.22 $\pm$ 0.06\\
		$\gamma$2 & 71.45 $\pm$ 0.05& 2.99 $\pm$ 0.11& 1.81 $\pm$ 0.02\\
		$\gamma$3 & 72.29 $\pm$ 0.13& 3.36 $\pm$ 0.35& 2.14 $\pm$ 0.07\\
		$\gamma$4 & 70.61 $\pm$ 0.20& 3.78 $\pm$ 0.55& 2.07 $\pm$ 0.11\\
		$\gamma$5 & 69.17 $\pm$ 0.75& 5.34 $\pm$ 2.16& 2.12 $\pm$ 0.30\\
		$\gamma$6 & -               & -              & -              \\
		\hline
	\end{tabular}
\end{table}



\section{Discussion}
\label{sec:discussion}

\begin{figure}
	\includegraphics[width=\columnwidth]{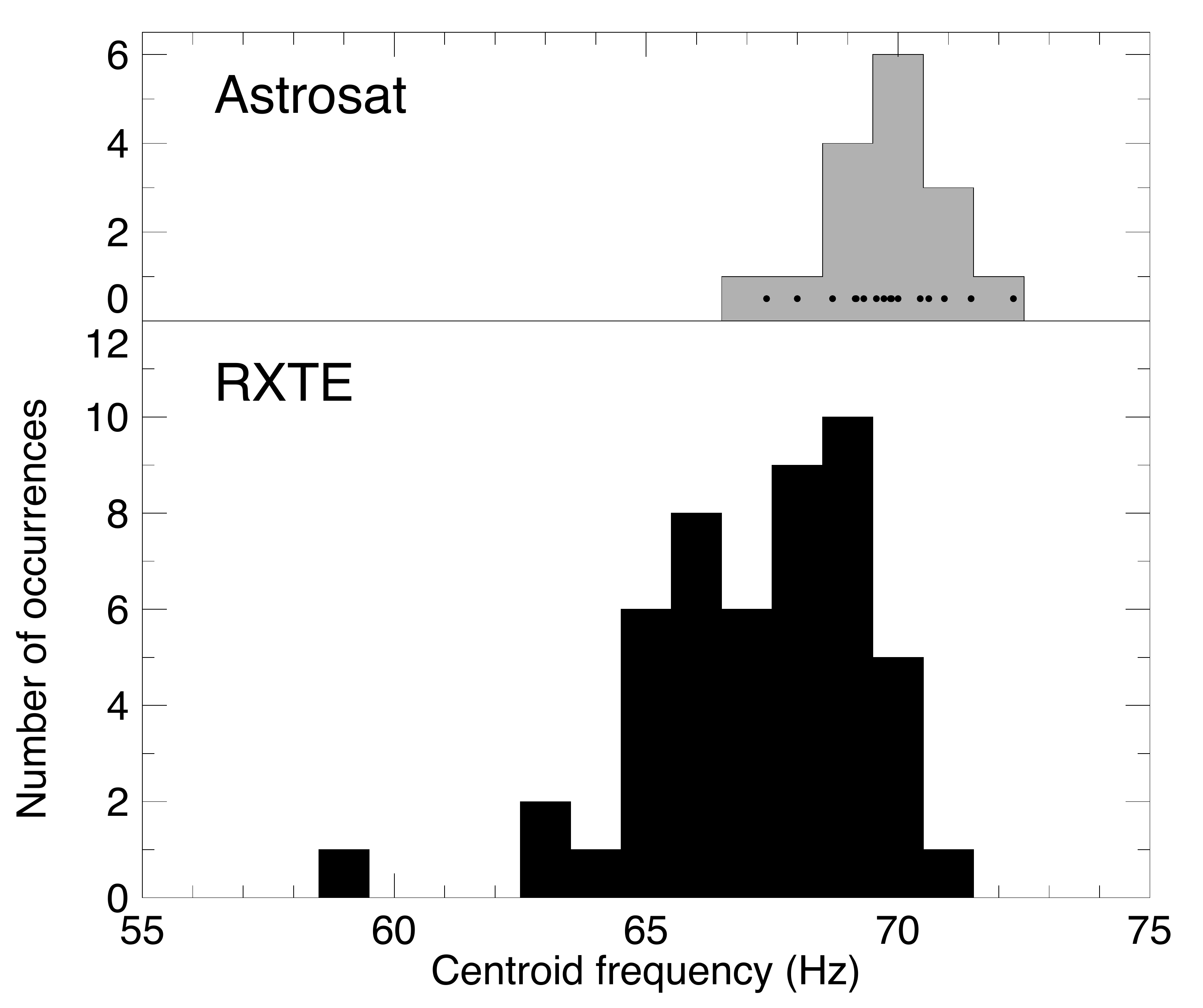}
    \caption{Top: distribution of the HFQPOs in Tab. \ref{tab:qpo}. The dots indicate the single values.
                  Bottom: distribution of the 51 HFQPOs detected with RXTE from Belloni \& Altamirano 2013a.
             }
    \label{fig:comparison}
\end{figure}

We have analyzed a set of observations of GRS 1915+105 obtained with the laxpc instrument on board \textsc{Astrosat} and found for the first time evidence for variations in the centroid frequency of the HFQPO, correlated with the position in the hardness-intensity diagram and therefore to spectral variations. Detailed spectral analysis from these data is difficult, given the complex selection of data points. Clearly, spectral variations along the CCD are strong (see Fig. \ref{fig:licu}), but at this stage our spectral analysis is still too uncertain and we will have to defer it to a future paper. 
The HFQPOs in GRS 1915+105 before \textsc{Astrosat} have been observed only with RXTE, from the initial discovery of Morgan et al. (1997). Additional peaks, simultaneous with the $\sim$70 Hz one and at different frequencies, have been detected previously (27 Hz: Belloni, M\'endez \& S\'anchez-Fern\'andez 2001; 41 Hz: Strohmayer 2001; 34 Hz: Belloni \& Altamirano 2013b). No additional peaks have been detected in our data.
Belloni \& Altamirano (2013a) have detected a HFQPO in 51 RXTE observations. Three of them belonged to variability class $\mu$ and nine to variability class $\omega$. In this work, we analyzed data that belong to these two classes, which see GRS 1915+105 reach the very particular region in the HID that corresponds to HFQPO detections (see Belloni \& Altamirano 2013a). Notice that the frequencies that we observe here are on the high side compared to those in Belloni \& Altamirano (2013a), where HFQPOs from class $\omega$ were also systematically higher in frequency. The distributions of frequencies for RXTE and this work are shown in Fig. \ref{fig:comparison}. This indicates that there are secular variations, but it is remarkable that at the distance of years the frequencies are still rather close.
The phase lags we detect are compatible in value with those reported by M\'endez et al. (2013), but we observe for the first time an evolution: higher region numbers have systematically lower hard lags. This is particularly true for the 20-30 keV vs. 5-10 keV lags, which decrease to the point of becoming negative for region 6. Contrary to centroid frequencies, the dependence on region is monotonic, indicating a more complex relationship with X-ray hardness. A full spectral analysis is required to link these changes to physical parameters in the accretion flow. 

\begin{figure}
	\includegraphics[width=\columnwidth]{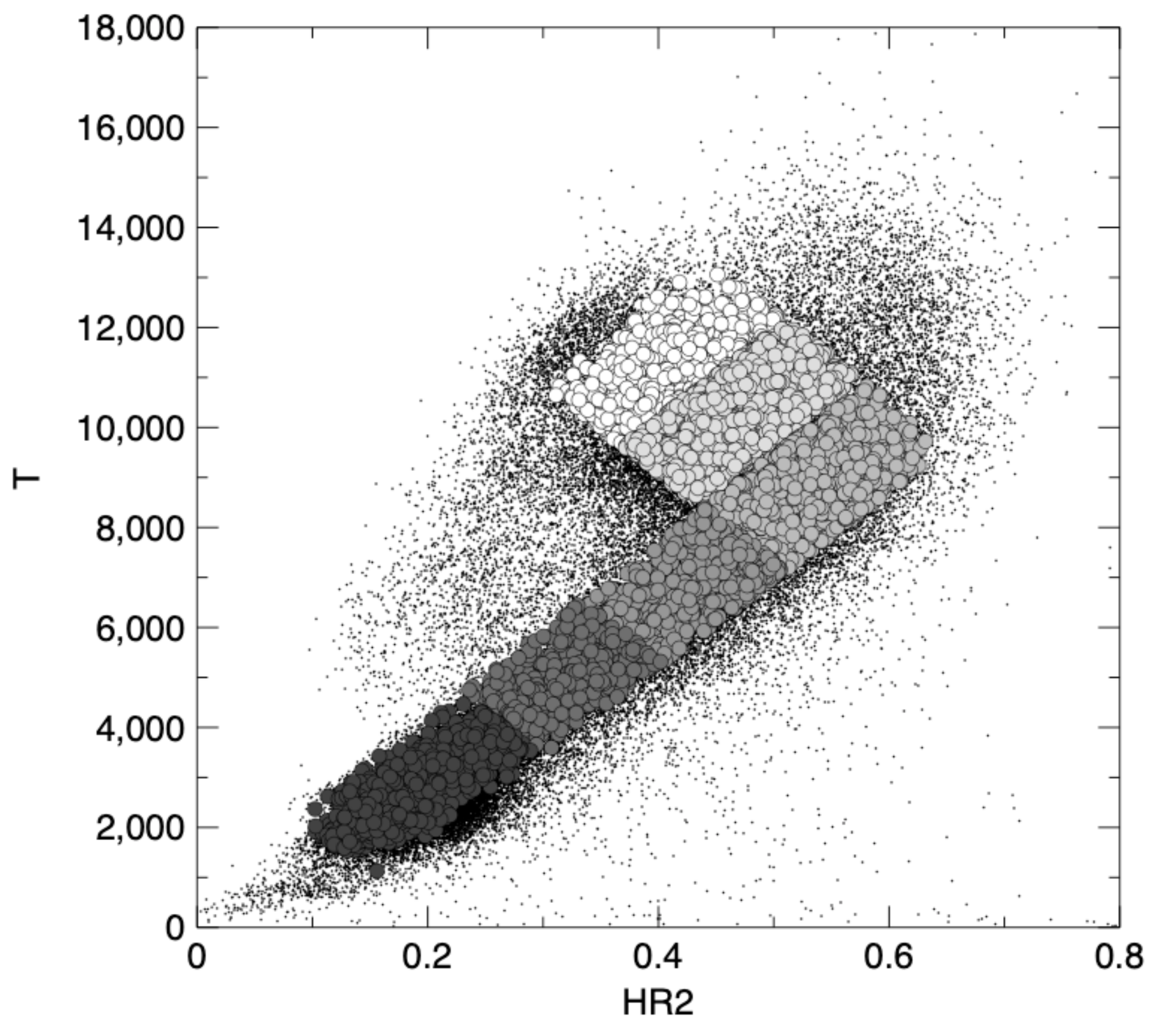}
    \caption{Original hardness-Intensity diagram for the points in group $\beta$ (black dots). The larger grayscale circles correspond to regions 1 (lighter) through 6 (darker).
             }
    \label{fig:beta_HID}
\end{figure}

HFQPOs are very elusive signals and few detections are available, all of them from the RXTE satellite (see Belloni, Sanna \& M\'endez 2012, Belloni \& Altamirano 2013a,b and references therein). Theoretical models have been put forward, but given the scarcity of data they cannot be tested against each other. The models have been extensively explored in the recent past. The relativistic-precession model (Stella \& Vietri 1998,1999) associates the high-frequency signals to either the Keplerian frequency or the periastron precession frequency at a certain radius of the accretion flow. If the radius at which the frequencies are produced varies, the frequency will vary correspondingly (see e.g. Motta et al. 2014a,b). The changes observed here are of the order of 6\%, which would correspond to a rather small change in radius, most likely not measurable with the current spectral uncertainties. However, we notice two things. The first is that with the current best mass estimate ($12.4^{+2.0}_{1.8}$M$_\odot$, Reid et al. 2014), even assuming a zero spin the lowest frequency reachable by a Keplerian frequency at the innermost stable circular orbit would be higher than 70 Hz. This means that this feature cannot be associated to that physical frequency at that radius (nor to the periastron precession frequency at the same radius, which has the same value). Of course it can be either of the two frequencies at a larger radius, which however would need to be rather stable throughout the years. 
The second is that our analysis shows that the HFQPO frequency increases with hardness, as the six regions identified in the H$'$I$'$D have the real measured hardness increasing from region 1 to 3, to decrease again to region 6 (see Fig. \ref{fig:beta_HID}, in which the HID for the $\beta$ group is shown, with the points from the six regions highlighted). Low-frequency QPOs, very common in black-hole binaries, always show a centroid frequency that decreases with hardness. Within the RPM, these low-frequency features are associated to the Lense-Thirring precession at the same radius at which the other two frequencies are produced and therefore are always positively correlated with them. An opposite correlation with spectral hardness does not point towards such a connection, although here no LFQPOs are detected and a direct comparison cannot be made. However, Yadav et al. (2018) have shown that in \textsc{Astrosat} observations of GRS 1915+105 during the $\chi$ variability class the LFQPO is positively correlated with hardness. The difference between this source and the other BHBs could be that the disk component in GRS 1915+105 is considerably hotter, leading to a significant disk contribution to the hard band. A comparison with the spectral parameters in future work will clarify this dependence.
An alternative model is the Epicyclic Resonance Model (Abramowicz and Kluzniak 2001), which associates the observed frequencies to relativistic frequencies at special radii when these are in resonance and therefore assume values in simple integer ratios. In this case we only observe one frequency, which prevents a measurement of a frequency ratio. We also observe 6\% variations in the centroid frequency, which in the model should remain constant as it is associated to a constant radius, determined solely by the mass and spin of the central black hole. However, the observed variations are small and it might be possible to reconcile them within the model.

\section{Conclusions}
We have measured for the first time variations in the centroid frequency of the HFQPO in GRS 1915+105, which was observed to vary between 67.4 and 72.3 Hz. The variations were observed to be correlated with the position on the hardness-intensity diagram, where both hardness and count rate varied by almost one order of magnitude. Systematic variations in the hard lags at the QPO frequency were also measured. Future work will deal with the challenging task of extracting and fitting energy spectra, but it is clear that the HFQPO frequency increases as the spectrum hardens, while no monotonic variation is observed with count rate, despite the large variation of the latter.

These results confirm that GRS 1915+105 is the best object available up to now to study HFQPOs, given the difficulty to observe them in other transients, both because of the faintness of the signal and its transient behaviour. The observed changes, while they cannot yet be applied to specific theoretical models, indicate that future timing missions with larger collective area such as eXTP will be able to follow in detail how the HFQPO varies as a function of spectral parameters and will shed light on the nature of these eslusive features.

\section*{Acknowledgements}

This work makes use of data from the \textsc{Astrosat} mission of the Indian Space Research Organisation (ISRO), archived at Indian Space Science Data Centre (ISSDC). This work has been supported by the Executive Programme for Scientific and Technological cooperation between the Italian Republic and the Republic of India for the years 2017-2019 under project IN17MO11 (INT/Italy/P-11/2016 (ER)). TMB acknowledges financial contribution from the agreement ASI-INAF n.2017-14-H.0. We thank an anonymous referee for his/her constructive comments.





\begin{thebibliography}{99}
\bibitem[\protect\citeauthoryear{}{1999}]{} Abramowicz M.A., Kluzniak W., 2001, A\&A, 374, L19
\bibitem[\protect\citeauthoryear{}{1999}]{} Agrawal, P.C., 2006, Adv. Sp. Res., 38, 2989
\bibitem[\protect\citeauthoryear{}{1999}]{} Antia H.M., Yadav J.S., Agrawal P.C., 2017, ApJS, 231, 10
\bibitem[\protect\citeauthoryear{}{1999}]{} Belloni T.M., M\'endez M., King A.R., et al., 1997a, ApJ, 479, L145
\bibitem[\protect\citeauthoryear{}{1999}]{} Belloni T.M., M\'endez M., King A.R., et al., 1997b, ApJ, 488, L109
\bibitem[\protect\citeauthoryear{}{1999}]{} Belloni T.M., Klein-Wolt M., M\'endez M., et al., 2000, A\&A, 355, 271
\bibitem[\protect\citeauthoryear{}{1999}]{} Belloni, T., M\'endez, M., \& S\'anchez-Fern\'andez, C. 2001, A\&A, 372, 551
\bibitem[\protect\citeauthoryear{}{1999}]{} Belloni T.M., 2010, in ``The Jet Paradigm'', T.M. Belloni ed., LNP, 794, 53
\bibitem[\protect\citeauthoryear{}{1999}]{} Belloni T.M., Sanna A., M\'endez M., 2012, MNRAS, 437, 2554
\bibitem[\protect\citeauthoryear{}{1999}]{} Belloni T.M., Altamirano D., 2013a, MNRAS, 432, 10
\bibitem[\protect\citeauthoryear{}{1999}]{} Belloni T.M., Altamirano D., 2013b, MNRAS, 432, 19
\bibitem[\protect\citeauthoryear{}{1999}]{} Fender R., Belloni T., 2004, Ann. Rev. Astr. Ap., 42,317
\bibitem[\protect\citeauthoryear{}{1999}]{} Hannikainen D.C., Rodriguez J., Vilhu O., et al., 2005, A\&A, 435, 995
\bibitem[\protect\citeauthoryear{}{1999}]{} Janiuk A., Czerny B., Siemiginowska A., 2000, ApJ, 542, L33
\bibitem[\protect\citeauthoryear{}{1999}]{} Klein-Wolt M., Fender R.P., Pooley G.P., et al., 2002, MNRAS, 331, 745
\bibitem[\protect\citeauthoryear{}{1999}]{} Leahy D.A., Darbro W., Elsner R.F., et al., 1983, ApJ, 266, 160
\bibitem[\protect\citeauthoryear{}{1999}]{} Morgan E.H., Remillard R.A., Greiner J., 1997, ApJ, 482, 1010
\bibitem[\protect\citeauthoryear{}{1999}]{} M\'endez, M., Altamirano, D., Belloni, T., Sanna, A., 2013, MNRAS, 432, 2132
\bibitem[\protect\citeauthoryear{}{1999}]{} Motta S.E., Belloni T.M., Stella L., et al., 2014a, MNRAS, 426, 1701
\bibitem[\protect\citeauthoryear{}{1999}]{} Motta S.E., Mu\~noz-Darias T., Sanna A., et al., 2014b, MNRAS, 439, L65
\bibitem[\protect\citeauthoryear{}{1999}]{} Reid M.J., McClintock J.E., Steiner J.F., et al., 2014, ApJ, 796, 2
\bibitem[\protect\citeauthoryear{}{1999}]{} Remillard R.A., Morgan E.H., McClintock J.E., et al., 1999, ApJ, 522, 397
\bibitem[\protect\citeauthoryear{}{1999}]{} Singh K.P., et al., 2014, in "Space Telescopes and Instrumentation 2014: Ultraviolet to Gamma Rays, p91441S, doi:10.1117/12.2062667
\bibitem[\protect\citeauthoryear{}{1999}]{} Stella L., Vietri M,, 1998, ApJ, 492, L59
\bibitem[\protect\citeauthoryear{}{1999}]{} Stella L., Vietri M,, 1999, Phys. Rev. Lett., 530, 350
\bibitem[\protect\citeauthoryear{}{1999}]{} Strohmayer, T.E., 2001, ApJ, 554, L169
\bibitem[\protect\citeauthoryear{}{1999}]{} Yadav J.S., et al., 2016, in "Space Telescopes and Instrumentation 2014: Ultraviolet to Gamma Rays, p99051D, doi:10.1117/12.2231857
\bibitem[\protect\citeauthoryear{}{1999}]{} Yadav J.S., Misra R., Verdhan Chauhan J., et al., 2018, ApJ, 833, 27
\end{thebibliography}




%
%


\bsp	
\label{lastpage}
\end{document}